\documentclass[aps,prl,twocolumn,amsmath,amssym,amsthm,superscriptaddress,reprint]{revtex4-2}
\pdfoutput=1
\usepackage[colorlinks=true,
linkcolor=blue,
urlcolor=blue,
citecolor=blue]{hyperref}
\usepackage{graphicx}
\usepackage{dcolumn}
\usepackage{color}
\usepackage{amsmath}
\usepackage{amssymb}
\usepackage[dvipsnames]{xcolor}
\usepackage[normalem]{ulem}

\begin{document}

\title{Physical Observers and Quantum Reconstructions}
\author{Dorian Daimer}
\author{Susanne Still}
\affiliation{Department of Physics and Astronomy \\ University 
of Hawaii at M\=anoa, Honolulu, HI 96822, USA}

\begin{abstract}
There is a multitude of interpretations of quantum mechanics, but foundational principles are lacking. Relational quantum mechanics views the observer as a physical system, which allows for an unambiguous interpretation as all axioms are purely operational, describing how observers acquire information. The approach, however, is based on the premise that the observer retains only predictive information about the observed system. 
Here, we justify this premise using the following principle: Physically embedded observers choose information processing strategies that provide them with the  option to approach physical limits to the greatest possible extent. Applied to a lower limit on energy dissipation, the principle leads directly to a compact predictive model, thus justifying this core premise of relational quantum mechanics.
\end{abstract}

\maketitle

Quantum theory \footnote{Here ``quantum mechanics'', ``quantum theory'' and ``quantum physics'' are used interchangeably.} has achieved unparalleled success in predicting and explaining a vast range of physical phenomena. However, even a century after its inception, no universally accepted interpretation of the theory exists. Textbook formulations of quantum mechanics rely on a set of highly abstract, mathematical axioms \cite{shankar2012principles, von2013mathematische, griffiths2019introduction}. 
Multiple interpretations co-exist, based on mutually contradictory concepts \cite{zeilinger1999foundational}. The best known is the Copenhagen interpretation \cite{Bohr1928, Heisenberg1958}, followed by alternatives such as Bohm's guiding wave interpretation \cite{bohm1952suggested}, the many worlds interpretation \cite{Everett1956}, the consistent histories interpretation \cite{Griffiths1984}, and the relational interpretation \cite{rovelli1996relational} to name just a few. It has been suggested, that the multitude of possible interpretations stems from quantum physics' lack of foundational principles \cite{rovelli1996relational, zeilinger1999foundational}. In response to this enduring challenge, many researchers have attempted to reconstruct quantum mechanics from different sets of axioms, leading to significant progress in various approaches such as quantum logic \cite{von1936logic, jauch1963can, isham1994quantum}, generalized probability theory \cite{hardy2001quantum, barnum2007generalized, barrett2007information, dakic2009quantum} and relational quantum mechanics \cite{rovelli1996relational}, though a complete and universally accepted reconstruction remains elusive. 

Relational approaches stand apart because they treat the observer as part of the theory, a physical system itself, which we will henceforth refer to as a ``physical observer". This is in contrast to a {\it deus ex machina}, an abstract observer imagined as outside of the scope of the physical theory \cite{baclawski2018observer, galili2024observer}. As a consequence of this different viewpoint, all the axioms in relational reconstructions are purely operational, describing how physical observers acquire and keep information. 

Importantly, the state of the observed system is assumed to be identical with a record of predictive information retained by the observer, that is, information enabling prediction of the system's behavior. This assumption is core to relational approaches \cite{rovelli1996relational, hohn2017toolbox, hohn2017quantum}.

Here we argue that this assumption can be derived from a principle governing how physical observers construct their recorded memory. We explain how compact predictive records emerge from choosing strategies that minimize a general lower bound on energy dissipation, placing a fundamental thermodynamic limit on the physical process necessary to record information. We discuss how this approach can be extended to articulate a principle for the emergence of physical observers, their behavior, and general rules for information acquisition, thus providing a foundational principle underpinning relational quantum mechanics. 

\section{Axiomatic Reconstructions of Quantum Theory}
Axiomatic reconstructions of quantum theory aim to identify the physical principles underlying its mathematical formalism, much like how special relativity follows from the principle of relativity and the invariance of the speed of light \cite{einstein1905elektrodynamik}. By clarifying these foundations, reconstructions can guide modifications of quantum theory \cite{goyal2010information, masanes2011derivation} and potentially point the way toward unification with gravity \cite{goyal2010information}.

Early reconstructions were based on mathematical axioms \cite{mackey1957quantum, pool1963simultaneous, ludwig2012foundations}, particularly through quantum logic, originally introduced to address the conceptual issues of textbook quantum mechanics \cite{von1936logic, dilworth1961lattice, redei1996john}. While quantum logic evolved into a rich discipline \cite{svozil1998quantum, coecke2000operational, wilce2002quantum}, it offered limited progress on foundational physics questions.

A major shift occurred around the early 2000s with the rise of instrumentalist and information-theoretic approaches. Hardy’s influential work \cite{hardy2001quantum} reframed quantum theory as a novel kind of probability theory, initiating a family of reconstructions within the framework of generalized probability theories \cite{barrett2007information, chiribella2011informational, janotta2014generalized, chiribella2015entanglement}, which include incompatible observables, using basic experimental concepts, preparation, transformation, and measurement, to represent physical systems operationally \cite{janotta2014generalized}. In parallel, foundational work by Rovelli \cite{rovelli1996relational} and others \cite{barnum2007generalized, fuchs2002quantum, clifton2003characterizing, grinbaum2003elements, brassard2005information, fivel2012derivation, masanes2013existence} emphasized the central role of information in quantum theory.

Reconstruction programs generally follow three steps: (1) propose physical principles to narrow the space of possible theories, (2) represent these principles mathematically, and (3) derive the formalism of the theory \cite{brukner2009information}, typically in the finite-dimensional case \cite{rovelli1996relational, hardy2001quantum}. Crucially, for reconstructions to clarify the {\it physical} content of quantum theory, the underlying principles must be operational, that is, based on experimentally accessible procedures and outcome \cite{zeilinger1999foundational}.

Instrumentalist approaches focus on laboratory contexts and experimental predictions, but the axioms themselves may remain abstract. In contrast, {\it operational} axioms are explicitly formulated in terms of physical procedures. While Hardy’s reconstruction is regarded by some as the first complete derivation of quantum mechanics from operational principles\cite{hohn2017quantum}, others argue that the postulates lack clear operational meaning \cite{brukner2009information, masanes2011derivation}. Later reconstructions addressed this by introducing more operationally grounded axioms, such as limits on information capacity \cite{brukner2009information}, the concept of an information unit \cite{masanes2013existence}, or using purification to single out quantum theory \cite{chiribella2011informational}.

Beyond generalized probability theories, other frameworks have been employed. For example an information-geometric approach using the Fisher–Rao metric \cite{goyal2010information}, inspired by a notion of statistical distance \cite{wootters1981statistical}. Other reconstructions impose information-theoretic constraints within a $C^*$-algebraic framework \cite{clifton2003characterizing}, or adopt abstract categorical formulations \cite{abramsky2004categorical, coecke2010quantum}.

All these diverse approaches have yielded deep insights into quantum theory’s mathematical structure, though not all equally illuminate its physical content. 

\subsection{Relational quantum mechanics and the physical observer}
A relational approach to quantum mechanics was first proposed by Rovelli, based on the observation that in quantum mechanics different observers can have a different yet valid description of the same sequence of events \cite{rovelli1996relational}. This is reminiscent of special relativity, where a previously absolute notion of time was replaced by a relative one \cite{einstein1905elektrodynamik}. 

In relational quantum mechanics the notion of observer-independent values of physical quantities is abandoned. Instead, all descriptions of physical systems are inherently context dependent, valid only with respect to a specific observer \cite{rovelli1996relational}. 

The relational view emphasizes that information is exchanged via physical interactions between the observer and the system. Formally these interactions can be modeled as a sequence of binary questions $(Q_1, Q_2, Q_3, ...)$ together with the binary answer string $(a_1, a_2, a_3, ...)$ where each $a_i \in \{0, 1\}$ represents the answer to question $Q_i$ \cite{rovelli1996relational}.

The potentially infinite string of answers, given the questions, contains all information the observer $O$ has about the system $S$. But not all of this information is necessary to make predictions about answers to future questions. 
Rovelli introduces a notion of {\it relevant} information as the subset of {\em predictive} answers 
\footnote{In this quote we replace Rovelli's notation for the answer string, $(e_1, e_2, e_3, ...)$, his equation (4), with our notation, $(a_1, a_2, a_3, ...)$, to improve readability.}: 
\begin{quote}
    The {\it relevant} information (from now on, simply information) that $O$ has about $S$ is defined as the non-trivial content of the (potentially infinite) string $(a_1, a_2, a_3, ...)$, that is the part of $(a_1, a_2, a_3, ...)$ relevant for predicting future answers to possible future questions. The relevant information is the subset of the string $(a_1, a_2, a_3, ...)$, obtained discarding the $a_i$'s that do not affect the outcomes of future questions \cite{rovelli1996relational}.
\end{quote}
This subset is then taken to be the observer-dependent system state. Rovelli stresses that the novelty is the observer-dependence \cite{rovelli1996relationalfootnote}. It is thus obvious that (i) the state has to be understood {\em conditional} on the questions the observer has asked, to which the answers were recorded, and (ii) the state is a condensed memory of the entire history of answers the observer recorded in response to questions.

Regarding point (i), it is important to note that the answer string by itself is meaningless. Only in conjunction with the questions do the bits encoded in the answer string attain any meaning. Regarding point (ii), it is implicitly assumed that the observer retains both questions and answers, because without that, the observer would not be able to make any predictions. 

The nomenclature uses {\it information} synonymously with {\it relevant information} to mean information relevant with respect to predicting the future behavior of the observed system. We call this {\it predictive information}.

A concrete mathematical framework together with an explicit reconstruction of $N$ qubit quantum mechanics was provided over 20 years later by H\"ohn \cite{hohn2017toolbox}. This approach shows that relational reconstructions can also be formulated based on the questions the observer asks \cite{hohn2017quantum}. Both views are equivalent. 
Thus, either questions or the respective answers can be used to describe systems in relational quantum mechanics. But in either case, the description is only meaningful if the observer has a record of {\it both} questions and answers. 
The system state is conceptually equated to what we shall henceforth call a {\em predictive record}. The set of questions leading to a predictive record is characteristic of a particular observer. 

All axioms required for relational reconstructions of quantum mechanics are based on this fundamental notion of the state of a quantum system as an observer-dependent predictive record.

\section{A case for prediction}
The underlying assumption---that it is the observer's role to {\em predict} the system's behavior---may appear self-evident to most physicists, as prediction lies at the heart of the scientific enterprise. But why should it be so? If we take the relational notion seriously that there is no fundamental distinction between the observer and the observed system, both are physical, then one must ask, why would there be a physical system {\em predicting} other system(s) it interacts with? Is this behavior inherent or emergent? Is there any physical reason for the emergence of such behavior? That is the core question we address here.

All processing of information---whether it involves acquisition, compression, or retention in memory---requires not only time and space but also energy when performed by a physical system. Fundamental bounds constrain the speed and compactness of such processing \cite{bennett1985fundamental, lloyd2000ultimate, levitin2009fundamental}, as well as the minimum amount of energy that must inevitably be dissipated in the process \cite{szilard1929german, parrondo2015thermodynamics, CB}.
How closely a physical observer can approach these fundamental bounds depends both on the observer’s strategy for information processing and on the specific physical implementation of that strategy. 

An information processing strategy determines what kind of record is kept in memory by specifying a rule for how to react in response to input data. Colloquially, there are many strategies which are so inefficient and wasteful that they preclude any approach to the fundamental limits, regardless of how they are implemented. We expect only a subset of strategies to support efficient performance. Similarly, we cannot expect all observer strategies to be equally effective at predicting the behavior of the observed system. As it turns out, these two properties---thermodynamic efficiency and predictive capability---are closely related \cite{still2012thermodynamics}.

To understand this, recall that all information processing requires an input. In the case of a physical observer, this input comes from a transient interaction with the observed system, causing changes in the state of the physical observer through coupling dynamics. In this sense, physical observers are driven systems—--driven by the change of some parameter or feature, as a function of the observed quantities--—and this driving requires energy, some of which may be dissipated.

Intuitively, the lower bound on dissipation should depend on {\em how much} information the observer chooses to retain in memory, as one would expect higher costs when more information is stored. But it turns out that it also depends on {\em what kind of} information is retained, and thereby on the observer's information processing strategy. For all driven systems, no matter how far they are driven from thermodynamic equilibrium, the lower bound on dissipation is proportional to the amount of non-predictive information the observer records about its environment \cite{still2012thermodynamics}. 

 Thus, retaining predictive information to the largest degree possible will result in a potential thermodynamic advantage by making the lower bound on dissipation smaller. Mathematically, the dependence of the bound on the strategy arises from the mutual information terms found in the bound \cite{still2012thermodynamics, CB}. This bound applies not only to classical systems but also to quantum systems \cite{grimsmo2016quantum}, providing a universal limit on the minimal dissipation encountered by physical observers, whether they are macroscopic classical systems or quantum systems. 

Whether this thermodynamic advantage is actually realized depends on the specific physical implementation of the memory retention process that executes the information processing strategy. 
Any given strategy would be realized through physical coupling and decoupling dynamics. This can, in principle, be implemented with a wide range of physical systems. While some implementations may achieve the strategy-dependent bound on dissipation, many others are likely to be suboptimal. In some cases, a single implementation with tunable parameters may realize the bound for certain parameter settings but not for others.
Moreover, physical constraints may impose trade-offs between competing objectives, for example, minimizing dissipation versus maximizing processing speed. A tunable implementation could allow the observer to adjust these trade-offs in a context-dependent manner, prioritizing one quantity over another depending on external conditions. This flexibility could be particularly advantageous for biological systems, where the relevant trade-offs vary with context: in some situations minimizing dissipation is essential \cite{yu2017energy, zhu2019energy, wang2025macroscopic}; in others, energy expenditure may be necessary to enable rapid or robust responses \cite{szabo2005light, demmig2006photoprotection, wilhelm2011energy, hall2024entropy}. Sometimes a quick reaction is critical for survival \cite{buskey2002escape, van2003escape, svetlichny2018swim}, whereas in other scenarios, time may not be a limiting factor \cite{oh2007thermodynamic, garcia2011thermodynamics}.

In any case, one can certainly argue that, absent additional considerations and constraints, it would not be justifiable for an observer to employ an information processing strategy that increases the lower bound on dissipation beyond what is  unavoidable. 
In fact, the thermodynamically rational choice is to use a strategy that minimizes the bound \cite{CB, still2022partially, daimer2023thermodynamically}. Such a strategy will necessarily correspond to the maximally predictive model achievable, given the total amount of information retained in the physical observer's memory. The specifics of the predictive model and the precise details of the record stored in memory will naturally depend on the observed system. 

\section{A simple example}
To make this connection between dissipation and prediction concrete in the context of reconstructions of qubit quantum mechanics, let us consider the simplest possible example. Assume that one spin $1/2$ system is observed by a macroscopic classical observer. 

Now consider any single projective measurement (a specific type of binary question), along any axis.
The answer to it, $\{ a_1 \}$, serves as the system's state in the relational interpretation. The observer's physical state, $s$, has to correspond to $\{ a_1 \}$. To record a binary answer, the observer memory has to have at least two distinct states. It suffices to retain one bit of information in memory. This bit is fully predictive, given knowledge of the chosen projective measurement (the question), as long as only this one question is asked. No non-predictive bits are retained in memory, and therefore 
%the energy used to set the memory is not inevitably lost. T
the lower bound on average dissipation is zero \cite{still2012thermodynamics, CB}.

In the original spirit of Rovelli's approach, the ``system state", given by $\{ a_1 \}$, and now, equivalently, represented by $s$, is entirely observer-dependent, and there can be as many fully predictive descriptions of the system as there are possible projective measurements (i.e., distinct ``questions" different observers can ask).

In contrast, assume the record is extended to keep answers from two different consecutive questions, $\{a_1, a_2 \}$. Two bits of information have to be recorded to memory, requiring at least four distinct states, $s$, for the physical observer. But, in the best case, there is one bit of predictive information. This is the case if the last question the observer obtained an answer to gets asked again. Without knowledge of the order in which the questions were asked, there is less predictive information in the record. 
%on average. 
For example, say both questions were to occur with equal probability, then there is half a predictive bit in the record. 
%The details are not that important, because a
As long as both questions are being asked, there will always be non-predictive information in the record, and therefore, the lower bound on dissipation 
%encountered
%irretrievably lost work required 
%in the physical process that implements the memory 
would be larger than zero. The argument easily extends to longer records, as more information is retained, but the amount of predictive information can never exceed one bit. Therefore, it would be thermodynamically rational to retain only one answer.

It is important to note that a degeneracy arises when we follow our argument for the emergence of a thermodynamically rational information acquisition rule. On the one hand, the optimally predictive strategy keeps one bit of information in memory, all of which is predictive. It therefore keeps zero bits of non-predictive information, leading to a lower bound of zero on average dissipation per interaction. On the other hand, doing nothing at all can also be achieved with zero dissipation. The difference is that doing nothing keeps zero bits about the observed system in memory and does not allow for any prediction. 

This degeneracy should be expected. If we take the premise of relational quantum mechanics seriously, that the observer is a physical system, then it is not surprising that there should be a degeneracy between systems that exhibit the emergent property of prediction and those that are not meaningfully correlated with any other system. After all, not all physical systems act as observers. By minimizing the strategy-dependent lower bound on dissipation, one can thus recover the emergence of both observer-like and non-observer-like physical systems as a consequence of an underlying physical principle.

\section{A working hypothesis to explore rules on information acquisition beyond prediction}
One of us has previously proposed that the basic idea, explored here in the context of energetic considerations, could be extended to other physical aspects of information processing, and might serve as a guiding principle \footnote{Discussed at a number of venues since 2014, e.g. \cite{StillTalk2019FQXiTuscany}}.
The proposal is that, in the absence of additional external constraints or considerations, physical observers would adopt data representation strategies that keep their options open with respect to the ability to approach fundamental physical limits. Choosing strategies that do not preserve this flexibility simply cannot be justified, unless other, extrinsic constraints or needs are present.

More precisely, the basic premise is that there are physical systems that have the capacity to implement a physical observer, by means of having the capacity to implement memory. Their physical interaction with the observed system determines an information processing strategy. 
The idea of a least self-impediment principle for physical observers is then that, in the absence of additional constraints and considerations:
\begin{center}
    Information processing strategies emerge in physical observers from allowing strategy-dependent bounds on physical quantities related to information processing to come as close as possible to the fundamental limits.
\end{center}
It is interesting to explore what kind of observer strategies emerge from this principle. We have investigated this in the context of dissipation \cite{CB, still2022partially, daimer2023physical, daimer2023thermodynamically}. The application to other physical limits, such as those on speed and accuracy, is the subject of ongoing research. 
 
It is important not to forget that the principle talks about extremizing {\em bounds} on physical quantities, rather than the physical quantities themselves. In that way, our proposal differs from previous ones put forth in neuroscience (e.g. \cite{barlow1989unsupervised, hasenstaub2010metabolic, bialek2012biophysics} and references therein), where it was proposed that extremiziation of certain physical quantities could serve as a principle underlying biophysical function. Systems with neurons certainly qualify as physical observers. However, whether or not it is an advantage or a disadvantage to extremize a physical quantity should be context dependent for biological systems. Indeed, when it comes to dissipation and entropy production there is evidence to that end, as in some contexts it makes sense to minimize dissipation \cite{glansdorff1973thermodynamic, prigogine1984order, sabater2006organisms, brown2019theory, sabater2022entropy, fujimoto2024game}, whereas in other contexts it does not \cite{martyushev2006maximum, perunov2016statistical, gnesotto2018broken, fang2020nonequilibrium, tabanera2025multiple}.

The proposal articulated here, to extremize strategy dependent bounds over all possible strategies, is different from the proposal to extremize physical quantities over all possible mechanisms. Our principle simply ensures that the information acquisition and processing strategies employed by the observer preserve the possibility of reaching meaningful physical limits, if necessary. This principle should  apply not only in the context of neuroscience, but broader, to any kind of physical system that has the potential to qualify as a physical observer, as primitive as it may be. 

\section{Conclusion}
At the core of relational quantum mechanics lies the premise that observers are physical systems themselves and that they predict the systems they observe. 
This letter argued that prediction {\em emerges} as a behavior of physical observers from an underlying principle, articulated here. Compact predictive records emerge from minimization of an information acquisition rule dependent lower bound on the average dissipation encountered when physically implementing the observer's memory. A simple example was given to illustrate the concept. 

We hope that the letter is food for thought for many researchers, and that it opens up novel directions in areas that explore real-world observers, such as neuroscience, biochemical networks, robotics and machine learning, as well as for the foundations of quantum mechanics. 

%TC:ignore
\section{Acknowledgments.}
\noindent We thank Rob Shaw and Munro Ferguson for helpful comments on the manuscript. This publication was made possible through the support of the ID\# 62312 grant from the John Templeton Foundation, as part of the \href{https://www.templeton.org/grant/the-quantum-information-structure-of-spacetime-qiss-second-phase}{‘The Quantum Information Structure of Spacetime’ Project (QISS)}. The opinions expressed in this publication are those of the authors and do not necessarily reflect the views of the John Templeton Foundation.
%TC:endignore

\bibliographystyle{unsrt}
\bibliography{observer}

\end{document}